\documentclass[fleqn,twoside]{article}
\usepackage{espcrc2-largecollab}
\usepackage{graphicx}
\usepackage{epsfig,xspace}

\newcommand{\diffunit}{s$^{-1}$cm$^{-2}$sr$^{-1}$GeV}
\newcommand{\pointunit}{s$^{-1}$cm$^{-2}$GeV}
\newcommand{\Nch}{$N_{\mathrm{ch}}$}
\newcommand{\ie}{{\it i.e.}}

\newcommand{\ea}{{\it et al.}}
\newcommand{\ic}{IceCube}

\begin{document}

\title{
\vskip  -1.2 cm
IceCube - the next generation neutrino telescope 
at the South Pole }

\author{
A. Karle for the IceCube Collaboration: \\
J.~Ahrens\address[MAINZ]{Institute of Physics, University of Mainz, Staudinger Weg 7, D-55099 Mainz, Germany},
J.N.~Bahcall\address[PRINCETON]{Institute for Advanced Study, Princeton, NJ 08540, USA}, 
X.~Bai\address[BARTOL]{Bartol Research Institute, University of Delaware, Newark, DE 19716, USA}, 
T.~Becka\addressmark[MAINZ], 
K.-H.~Becker\address[WUPPERTAL]{Fachbereich 8 Physik, BUGH Wuppertal, D-42097 Wuppertal, Germany},
D.Z.~Besson\address[KANSAS]{University of Kansas, Lawrence, KS, USA }
D.~Berley\address[MARYLAND]{Dept. of Physics, University of Maryland, College Park, MD 20742, USA},
E.~Bernardini\address[DESY]{DESY-Zeuthen, D-15735 Zeuthen, Germany},
D.~Bertrand\address[ULB]{Universit\'e Libre de Bruxelles, Science Faculty CP230, B-1050 Brussels, Belgium}, 
F.~Binon\addressmark[ULB], 
A.~Biron\addressmark[DESY],
S.~B\"oser\addressmark[DESY], 
C.~Bohm\addressmark[STOCKHOLM],
O.~Botner\address[UPPSALA]{Division of High Energy Physics, Uppsala University, S-75121 Uppsala, Sweden},
O.~Bouhali\addressmark[ULB], 
Th.~Burgess\address[STOCKHOLM]{Dept. of Physics, Stockholm University, SCFAB, SE-10691 Stockholm, Sweden},
T.~Castermans\address[MONS]{University of Mons-Hainaut, 7000 Mons, Belgium},
D.~Chirkin\address[BERKELEY]{Dept. of Physics, University of California, Berkeley, CA 94720, USA},
J.~Conrad\addressmark[UPPSALA], 
J.~Cooley\address[MADISON]{Dept. of Physics, University of Wisconsin, Madison, WI 53706, USA},
D.F.~Cowen\address[PSU]{Dept. of Physics, Pennsylvania State University, University Park, PA 16802, USA},
A.~Davour\addressmark[UPPSALA], 
C.~De~Clercq\address[VUB]{Vrije Universiteit Brussel, Dienst ELEM, B-1050 Brussels, Belgium},
T.~DeYoung\addressmark[MADISON]\thanks{Present addr.: Santa Cruz Institute for Particle Physics, Univ. of Cal., Santa Cruz, CA 95064, USA.}, 
P.~Desiati\addressmark[MADISON], 
J.-P.~Dewulf\addressmark[ULB], 
B.~Dingus\addressmark[MADISON],
R.~Ellsworth\addressmark[MARYLAND],
P.A.~Evenson\addressmark[BARTOL],
A.R.~Fazely\address[SOUTHERN]{Dept. of Physics, Southern University and A\&M College, Baton Rouge, LA 70813, USA},
T.~Feser\addressmark[MAINZ], 
T.K.~Gaisser\addressmark[BARTOL], 
J.~Gallagher\addressmark[UW-ASTRO],
R.~Ganugapati\addressmark[MADISON],
A.~Goldschmidt\address[LBNL]{Lawrence Berkeley National Laboratory, Berkeley, CA 94720, USA},
J.~Goodman\addressmark[MARYLAND],
A.~Hallgren\addressmark[UPPSALA], 
F.~Halzen\addressmark[MADISON], 
K.~Hanson\addressmark[MADISON], 
R.~Hardtke\addressmark[MADISON], 
T.~Hauschildt\addressmark[DESY], 
M.~Hellwig\addressmark[MAINZ], 
P.~Herquet\addressmark[MONS],
G.C.~Hill\addressmark[MADISON], 
P.O.~Hulth\addressmark[STOCKHOLM], 
K.~Hultqvist\addressmark[STOCKHOLM], 
S.~Hundertmark\addressmark[STOCKHOLM], 
J.~Jacobsen\addressmark[LBNL], 
G.S.~Japaridze\address[ATLANTA]{CTSPS, Clark-Atlanta University, Atlanta, GA 30314, USA}, 
A.~Karle\addressmark[MADISON], 
L.~K\"opke\addressmark[MAINZ], 
M.~Kowalski\addressmark[DESY], 
J.I.~Lamoureux\addressmark[LBNL], 
H.~Leich\addressmark[DESY], 
M.~Leuthold\addressmark[DESY], 
P.~Lindahl\addressmark[KALMAR], 
I.~Liubarsky\address[IMPERIAL]{Imperial College London, Exhibition Road, London SW7 2AZ, UK}
J.~Madsen\address[UWRF]{Physics Dept., University of Wisconsin, River Falls, WI 54022, USA},
P.~Marciniewski\addressmark[UPPSALA], 
H.S.~Matis\addressmark[LBNL], 
C.P.~McParland\addressmark[LBNL], 
Y.~Minaeva\addressmark[STOCKHOLM], 
P.~Mio\v{c}inovi\'c\addressmark[BERKELEY], 
R.~Morse\addressmark[MADISON], 
R.~Nahnhauer\addressmark[DESY],
T.~Neunh\"offer\addressmark[MAINZ], 
P.~Niessen\addressmark[VUB], 
D.R.~Nygren\addressmark[LBNL], 
H.~Ogelman\addressmark[MADISON], 
Ph.~Olbrechts\addressmark[VUB], 
C.~P\'erez~de~los~Heros\addressmark[UPPSALA], 
A.C.~Pohl\addressmark[KALMAR], 
P.B.~Price\addressmark[BERKELEY], 
G.T.~Przybylski\addressmark[LBNL], 
K.~Rawlins\addressmark[MADISON], 
E.~Resconi\addressmark[DESY], 
W.~Rhode\addressmark[WUPPERTAL], 
M.~Ribordy\addressmark[DESY], 
S.~Richter\addressmark[MADISON], 
H.-G.~Sander\addressmark[MAINZ], 
T.~Schmidt\addressmark[DESY], 
D.~Schneider\addressmark[MADISON], 
D.~Seckel\addressmark[BARTOL], 
M.~Solarz\addressmark[BERKELEY], 
L.~Sparke\addressmark[UW-ASTRO],
G.M.~Spiczak\addressmark[UWRF], 
C.~Spiering\addressmark[DESY], 
T.~Stanev\addressmark[BARTOL], 
D.~Steele\addressmark[MADISON], 
P.~Steffen\addressmark[DESY], 
R.G.~Stokstad\addressmark[LBNL], 
P.~Sudhoff\addressmark[DESY],
K.-H.~Sulanke\addressmark[DESY], 
G.W.~Sullivan\addressmark[MARYLAND],
T.~Sumners\addressmark[IMPERIAL]
I.~Taboada\address{Departamento de F\'{\i}sica, Universidad Sim\'on Bol\'{\i}var, Apdo. Postal 89000, Caracas, Venezuela}, 
L.~Thollander\addressmark[STOCKHOLM], 
S.~Tilav\addressmark[BARTOL], 
C.~Walck\addressmark[STOCKHOLM], 
C.~Weinheimer\addressmark[MAINZ], 
C.H.~Wiebusch\addressmark[DESY]\thanks{Present address: CERN, CH-1211, Gen\`eve 23, Switzerland.}
Ch.~Wiedemann\addressmark[STOCKHOLM],
R.~Wischnewski\addressmark[DESY], 
H.~Wissing\addressmark[DESY], 
K.~Woschnagg\addressmark[BERKELEY],
Sh.~Yoshida\address[CHIBA]{Dept. of Physics, Faculty of Science, Chiba University, Chiba 263-8522, Japan}
}

\clearpage

\begin{abstract}
IceCube is a large neutrino telescope of the next generation to be
constructed in the Antarctic Ice Sheet near the South Pole.  We present
the conceptual design and the sensitivity of the IceCube detector to
predicted fluxes of neutrinos, both atmospheric and extra-terrestrial.  A
complete simulation of the detector design has been used to study 
the detector's capability to search for neutrinos from
sources such as active galaxies, and gamma-ray bursts.
\end{abstract}

\maketitle

\section{INTRODUCTION}

\begin{figure}[htp]
\centering
\epsfig{file=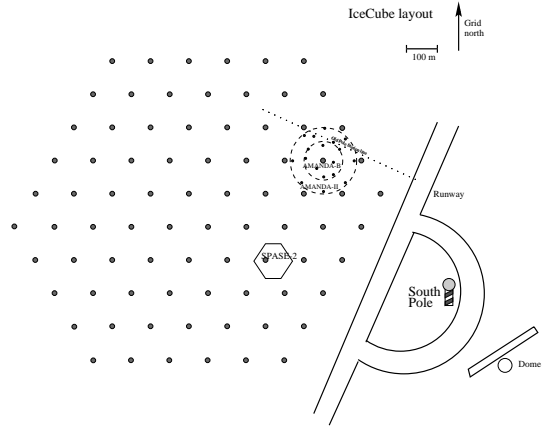,width=0.45\textwidth}
\vskip -0.4cm
\caption[1]{
\label{fig:geometry}Schematic diagram of the arrangement of the strings of the
\ic~
detector at the South Pole station. The existing AMANDA detector and the SPASE 
air shower array will be embedded in the IceCube array. 
}
\vskip -0.6 cm
\end{figure}

The successful deployment and operation of the AMANDA detector have 
shown that the Antarctic Ice sheet is an ideal medium and location 
for a large neutrino telescope. The detection of atmospheric neutrinos 
in agreement with expectations \cite{nature,atmospheric} established 
AMANDA as a neutrino telescope. Searches for neutrinos from 
Supernova \cite{sn}, dark matter \cite{dark}, point sources of muon 
neutrinos \cite{point} and 
diffuse sources  of high energy electron \cite{electron} and muon 
neutrinos \cite{difficrc} have demonstrated the physics potential of a deep ice 
neutrino detector.  However, a much larger detector is needed to reach a
sensitivity required for the detection of many predicted neutrino fluxes. 
Current and proposed detectors at sites in the northern hemisphere are discussed in
\cite{other-experiments}. 
IceCube is a projected under ice neutrino detector consisting of 4800 PMT
on 80 strings distributed over an area of 1\,km$^{2}$ (figure
\ref{fig:geometry})
and instrumented at a depth between 1400\,m and 2400\,m.
Detailed documentation can be found in  \cite{proposal-pdd}. 
A surface airshower detector consisting of 160 stations 
over 1\,km$^{2}$ augments the deep ice component by providing
a tool for calibration, background rejection and air shower physics.
The Monte-Carlo simulation
tools used to determine and optimize the performance of IceCube 
are verified with AMANDA data. They correctly describe the 
cosmic ray muon flux and the atmospheric neutrino flux \cite{atmospheric}. 

One of the principal objectives of IceCube is the detection of sources of
high energy neutrinos of astrophysical origin.  IceCube is sensitive to all
neutrino flavors over a wide range of energies.  Muons can be observed from
about $10^{11} eV $ to $10^{18} eV$ and beyond.  Cascades,
generated by $\nu_e$, $\overline{\nu}_e$, $\nu_\tau$, and
$\overline{\nu}_\tau$ can be observed and reconstructed at energies above
$10^{13} eV$.  Tau events can be identified above energies of about a PeV.
Interaction with the Earth will modulate the neutrino fluxes emerging at an
underground detector.

The potential backgrounds of large underground detectors are  downgoing
cosmic ray muons, atmospheric neutrinos, and the dark noise
signals detected in photomultipliers.  The simulated trigger  rate of 
downgoing cosmic ray muons in IceCube is 1700\,Hz while the rate
of atmospheric neutrinos ($\nu_\mu$ and $\overline{\nu}_\mu$) at trigger
level is about 300/day.  Depending on the type of signal to be searched for
this background is rejected using direction, energy, and neutrino flavor. 
At energies below $\approx$1\,PeV, neutrino astronomy must focus on upward going
neutrinos.  At energies above 1 PeV, the cosmic ray muon background
disappears while the low energy cosmic ray background can be rejected 
using an energy rejection cut.  The background contribution 
from dark noise is not significant because the total dark noise rate of the optical sensor
in situ is expected to be less than 0.5\,kHz.

The highest sensitivity for astrophysical point sources can be achieved 
with muons. The muon channel stands out for two reasons.
1.) Muons allow a very good angular resolution of $Å0.7^\circ$ over a wide energy range. 
2.) The effective volume for muons exceeds the geometric 
volume of the detector by factors of 10 to more than 50 depending on 
energy.
Due to the long range of high energy muons the interaction of the $\nu_\mu$ 
can take place at distances of tens of kilometers outside of the detector. 
We will focus on the muon detection because it provides the benchmark 
sensitivity for some of the fundamental goals of high energy neutrino 
astronomy.

\section  {DETECTOR DESIGN}

The detector consists of 4800 photomultipliers arranged on 80 strings 
at depths of 1400 to 2400\,m. The strings are arranged in a regular spaced
grid covering a surface area of 1\,km$^2$ as shown in figure 
\ref{fig:geometry}.
A surface component of 160 ice Cherenkov tanks provides a detector for air
showers with an energy threshold of about 1 PeV. Each string consists of a
60 optical modules (OM) spaced at 17\,m.  The geometric arrangement and the
total number of OM and strings is a result of MonteCarlo simulations with a
variety of different geometries.  The photomultiplier signals will be
digitized inside the pressure housing \cite{stokstad}.  The digitized
signals are given a global time stamp with a precision of $<$5\,ns and
transmitted to the surface.  The digital messages are sent to a string
processer, a global event trigger and event builder.  All time calibrations
will be automated.  The geometry of the detector will be known to a
precision of better than 2\,m initially after the deployment.  It will be
calibrated more precisely ($<1$\,m) with light flashers on board the OM and
with cosmic ray muons.  Both methods have been applied successfully in
AMANDA. High energy signals and complex events can be calibrated with
powerful lasers deployed with the detector.  The absolute orientation can
be calibrated with coincidences with the surface air shower component,
IceTop, and over longer time scales, with the observations of the
shadow of the moon.  Once events are built in the surface DAQ, data 
will be
processed and filtered. A reduced data set will be sent to the Northern
hemisphere for further processing and data analysis on a daily basis. 
Twenty four
hour satellite connectivity will be available for important messages. 
Construction of the detector is expected to commence in the Austral summer
of 2004/2005 and continue for 6 years.  The growing detector will be in
operation during construction, with each string coming online within days
after deployment.

\section{FUNDAMENTAL PERFORMANCE PARAMETERS}
\label{sec:performance}

In detailed simulations we compared the response of the detector to cosmic
ray muons, to atmospheric neutrinos and to a hypothetical hard neutrino
spectrum as generated by a shock acceleration mechanism 
\cite{performance}.  The event rates
generated by all three fluxes, normalized to one year of on-time, are
listed in table~\ref{tab:passclasses}.  
The event rates are given at trigger level (minimum of 5 OM signals 
within a section of a string) and for full event reconstruction with cuts applied for the
rejection of the cosmic ray muon background, which we refer to as ``level\,2''. 
Unless specified otherwise, basic background rejection cuts (level 2) are 
applied for all results shown.
 With an assumed flux strength of $E_{\nu}^{2} \times dN_{\nu}/dE_{\nu}$ =
10$^{-7}$\,\diffunit~for cosmic neutrinos, a factor of 10 below the current
best limit to high energy diffuse neutrinos \cite{difficrc}, we expect more
than 1000 signal events.  At this stage,
both, the background from atmospheric neutrinos and the background from
cosmic ray muons amount to roughly $10^{5}$ events
atmospheric neutrino induced muon events 
is based on \cite{lipariatm}) , including about 5000
events attributed to
prompt decays of charmed mesons \cite{rqpm}.  The application of an energy cut will
quickly suppress remaining cosmic ray muon background.

\begin{table}[htb]
\vskip -0.6cm
{\footnotesize
\renewcommand{\arraystretch}{1.2}
  \begin{center}
   \begin{tabular}{|c||c|c|c|}
    \hline
    ~           &  Trigger                                       &   Level 2    \\
    \hline \hline
    {\bf Cosmic $\nu$}    &  3.3$\times 10^{3}$      &  1.1 $\times 10^{3}$ \\
    \hline
  {\bf Atm $\nu$}    &  8.2  $\times 10^{5}$                &  9.6  $\times 10^{4}$    \\
    \hline
    {\bf Atmosph. $\mu$} & 4.1 $\times 10^{10}$  &  10  $\times 10^{4}$       \\ 
     \hline
    \end{tabular}
    \caption{Event rates are given for 1 year of signal and background.
The signal expectation corresponds to 
a source flux of $ E^{2}_{\nu} \times dN_{\nu}/dE_{\nu}$ = 10$^{-7}$\,\diffunit. 
The expectation for atmospheric neutrino induced muon events 
is based on \cite{lipariatm}) and it includes the 
prompt component according to \cite{rqpm}(rqpm). }
\label{tab:passclasses}
\end{center}
}
\end{table}

\begin{figure}[htp]
\begin{center}
\epsfig{file=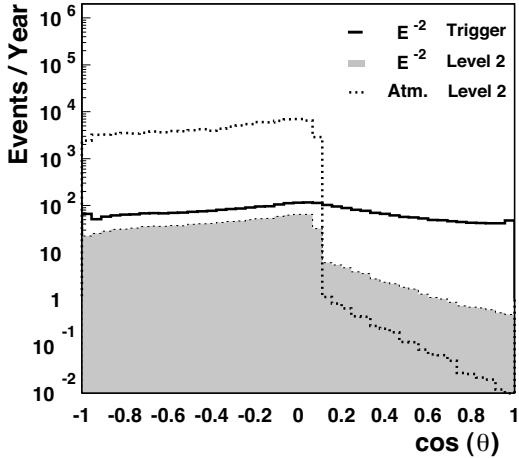,
width=.45\textwidth
}
\end{center}
\vskip -1.0cm
\caption{Distribution of the reconstructed zenith angle
a) Signal from a $E^{-2}$-source  at trigger level, 
b) with background rejection (level 2, shaded), and c) atmospheric neutrino 
background at level 2.
Event numbers are normalized to one year.
\label{fig:zenrej}
}
\end{figure}
Figure \ref{fig:zenrej} shows reconstructed zenith angle distributions 
of atmospheric neutrinos and of the injected cosmic neutrino flux.
Downgoing atmospheric neutrinos are highly suppressed at 
the level because low energy cosmic ray muons need to be 
rejected with energy cuts. The zenith angle  
distribution of the astrophysical neutrino spectrum 
is modified little  below the horizon. Above the horizon
the losses are much smaller than for atmospheric neutrinos
because of the harder spectrum. 
However, no energy cut has yet been applied yet to upgoing 
muons. 

\subsection{Effective Detector Area}
The detector sensitivity can be expressed in several quantities.
One possible choice is the ``effective detector area''.  
It is defined as 
\begin{equation}
    A_{\mathrm{eff}}(E_{\mu},\Theta_{\mu} ) =
  \frac{N_{\mathrm{detected}}(E_{\mu}, \Theta_{\mu})}{N_{\mathrm{generated}}(E_{\mu},\Theta_{\mu} )} \times A_{\mathrm{gen}},
\end{equation}
where
$N_{\mathrm{generated}}$ is the number of muons in the test sample that have 
an energy
$E_\mu$ at any point within the fiducial volume and an 
incident zenith angle $\Theta_{\mu}$. We take the point of closest approach 
to the detector center (which might also lie outside the geometrical
detector volume) for the energy dependency.
$N_{\mathrm{detected}}$ is 
the number of such events that are triggered or pass the cut level
under consideration.

\begin{figure}[htp]
\begin{center}
\vskip -0.8 cm
\epsfig{file=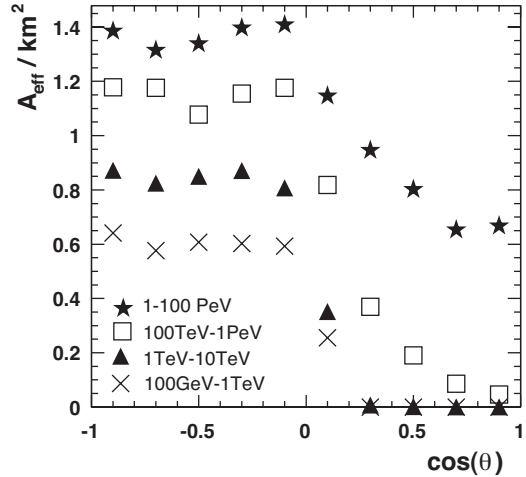,
width=.45\textwidth
}
\end{center}
\vskip -1.0cm
\caption{The effective area is shown as a function of the zenith angle 
after applying Level\,2 cuts for four ranges of energy.
\label{fig:aeff.en}
}
\end{figure}

The effective trigger area reaches one square kilometer at an energy of a
few hundred GeV. Roughly 50\% of all triggered events pass the ``standard
selection'' (Level\,2), independent of the muon energy.  Figure
\ref{fig:aeff.en} shows the effective areas for four separate energy ranges
as a function of the zenith angle of the incident muon
tracks.  The detector will provide an effective detection area of one
square kilometer for upward moving muons in the TeV range.  Above 100\,TeV
the selection allows for a detection of downgoing neutrinos,
\ie~ for an observation of the southern  
hemisphere ($\cos \theta > 0$). In the PeV range the effective area
for downgoing muons is above 0.6 km$^{2}$, increasing towards the horizon.
This means that \ic~ can be operated as a full sky observatory for 
PeV to EeV neutrinos.

\subsubsection{Angular resolution}

\begin{figure}[htp]
\begin{center}
\epsfig{file=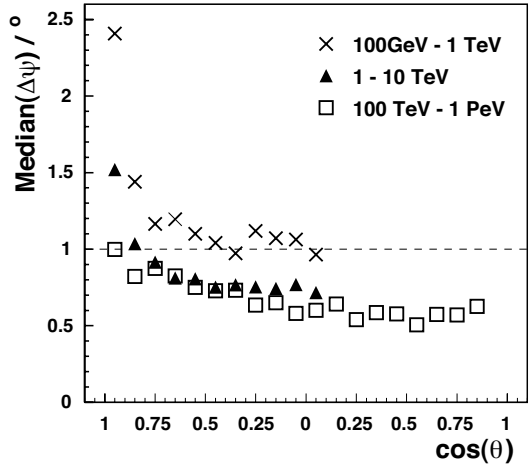,
width=.45\textwidth
}
\end{center}
\vskip -1.0cm
\caption{ Pointing resolution for muons.
Shown is the median space angle error of the reconstructed direction
as a function 
of  the zenith angle of the incident track. 
\label{fig:pointing}
}
\end{figure}
The angular resolution of the detector is an important quantity for the search 
for  neutrinos from point sources. Using the angular resolution of the 
detector,
background events can be more easily eliminated by restricting the search 
to a small angular region about the known direction of the object under 
investigation.
The median angular resolution (median pointing error) at different zenith angles is shown
 in figure \ref{fig:pointing}.

The angular resolution is about $0.7^\circ$ at TeV energies
and approaches $0.6^\circ$ near the horizon.    
We expect significant improvement of the pointing resolution with further 
development of the
reconstruction algorithms, in particular from including amplitude or 
waveform information.

\section{SENSITIVITY TO ASTROPHYSICAL NEUTRINO FLUXES}

\subsection{Diffuse fluxes}
Theoretical models of astrophysical fluxes of high energy neutrinos 
are often linked to the known flux of very high energy cosmic rays, 
which are believed to be of extragalactic origin, see for example 
\cite{uhe-nu-astro,WBUB,MPR98,SS96}. 
Many models have been developed that predict the expected flux of
neutrinos from the sum of all active galaxies in the universe.
First we will consider the sensitivity of \ic~ to 
detect a generic diffuse flux following an $E^{-2}$ spectrum.
The harder energy spectrum of an astrophysical neutrino flux 
can be used to discriminate the atmospheric
neutrino spectrum from the signal. 
We use a very simple and robust observable, the number of 
optical modules which have seen at least one photon, called the channel
multiplicity ($N_{\mathrm{ch}}$), as an energy separation cut.
Atmospheric neutrino induced events have typical 
channel multiplicities (event sizes) of about 30 to 60 channels.
The assumed high energy signal flux dominates the atmospheric neutrino 
background above channel multiplicities of about 200.

\begin{figure}[htp]
\vskip -0.0 cm
\begin{center}
\epsfig{file=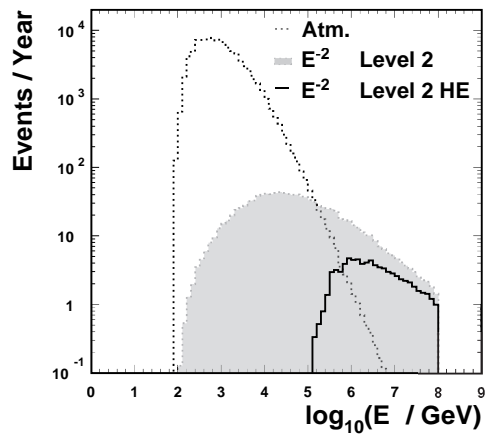,
width=.42\textwidth
}
\end{center}
\vskip -1.0cm
\caption{ Energy spectra of detected neutrinos for a hypothetical 
$E^{-2}$
source before (shaded) and after energy cut  (solid line), 
 and atmospheric neutrinos before the energy cut (dashed). 
\label{fig:nu.energy.spectra}
}
\end{figure}

The simulated source strength of $E_{\nu}^{2} \times dN_{\nu}/dE_{\nu} = 1
\times 10^{-7}$\,\diffunit~ would result in an expectation of 74 signal
events in one year of operation, compared to 8 background events from
atmospheric neutrinos, after applying an optimal energy cut 
($N_{\mathrm{ch}}>227$).  The background
expectation was calculated using the prompt charm prediction 
of Bugaev \ea \cite{rqpm},
according to which, prompt decays of charmed mesons 
contribute 80\% to the final atmospheric sample.
After three years of operation an overall flux limit of 
 $E^2 \times dN_{\nu}/dE_{\nu} = 4.2 \times 10^{-9}$\,\diffunit is 
 obtained. This is more than 
two orders of magnitude below the current limit obtained with AMANDA. 
This sensitivity would improve by a factor of 2 when using 
the charm prediction of TIG \cite{TIG}.
It should be noted that the sensitivity for diffuse 
 $\nu_e$- and $\nu_\tau$-fluxes 
may improve the overall sensitivity significantly. 
The energy spectra of the incident signal and background neutrinos  
are shown in figure \ref{fig:nu.energy.spectra}. The applied energy cut
results in an detection threshold of about 200\,TeV.
The sensitivity  obtained after one year of data taking
is already well below the Waxmann and Bahcall diffuse bound 
\cite{W&B}.

\begin{figure}[htp]
\begin{center}
\epsfig{file=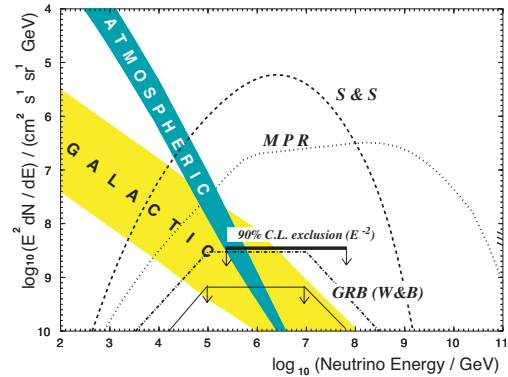,width=0.45\textwidth
}
\end{center}
\vskip -1.1 cm
\caption[1]{ Expected sensitivity of the \ic~ detector. The thick solid 
line indicates the 90\%\,c.l. limit setting potential for an E$^{-2}$
type spectrum for a time period of three years (see text for details).  
The thin solid line illustrates the GRB sensitivity for the reference 
spectrum \cite{WBUB}. 
\label{fig:intensities} 
}
\end{figure}

Apart from the generic case of a $E^{-2}$ spectrum, which is typical for 
scenarios that involve meson production in interactions of Fermi 
accelerated 
cosmic rays with matter,
we give the sensitivity to two other examples of diffuse models. 
Mannheim, Protheroe and Rachen have calculated an upper bound on the 
diffuse
neutrino flux arising from photohadronic interactions in unresolved
AGN jets in the universe. Their flux bound is shown in figure 
\ref{fig:intensities} labeled {\bf \emph{MPR}}, as a function of energy. 
Also shown is a model by Stecker and Salamon 
\cite{SS96} for
proton interactions on the UV thermal photon field in AGN cores, 
labeled {\bf \emph S\&\emph S}. 
When testing the sensitivity of IceCube to these specific models we found that 
IceCube would be sensitive to a flux of 2\% of the {\bf \emph{MPR}} 
flux, and $2.3 \cdot 10^{-3}$ of the Stecker/Salamon flux.
Also shown is the GRB flux prediction by Waxmann and Bahcall \cite{WBUB} 
(dash-dotted line). Here, the
sensitivity limit, defined as a potential exclusion at 90\% C.L., 
was calculated for an observation of 500 bursts
(expectation for one year) in the northern sky (see
section \ref{sec:grb}).

\vskip -1.0cm

\subsection{Sensitivity to Point Sources}
\label{sec:point.sens}
An excess of events  from a particular direction in the sky indicates the 
existence of a point source. The angular resolution of $0.7^\circ$ 
 allows using a search window of one degree radius that greatly reduces the background, 
while retaining a large fraction of the signal. 
In order to reject the remaining small number of atmospheric neutrinos 
we add a soft energy cut by requiring a channel multiplicity of 
\Nch=30.
The applied energy cut suppresses atmospheric neutrinos of energies mostly 
below 1\,TeV allowing a essentially background free 
detection potential for point sources. 
With these parameters we find an average flux upper limit of 
$E_{\nu}^2 \times  dN_{\nu}/dE_{\nu} = 5.5 \times 10^{-9}$\,\pointunit~
after one year of data taking. 
After five years of operation the sensitivity level would reach
$E^{2} \times dN_{\nu}/dE_{\nu} \sim 1.7 \times 10^{-9}$\,\pointunit~.

\subsection{Gamma Ray Burst Sensitivity}
\label{sec:grb}
The origin of the high-energy gamma-ray burst emissions seen by orbiting
satellites is still undetermined.  If the acceleration
mechanisms in these objects involve hadronic processes then the observed
fluxes of gamma-rays should be accompanied by neutrinos.  The detection of
these neutrinos is made easier by using the time stamp and direction provided by the
satellite observation.  This small duration of gamma ray bursts allows for
an essentially background-free search opportunity.
Waxman and Bahcall \cite{WBUB} calculated the expected flux of neutrinos
 from the sum of all gamma-ray bursts by assuming that the acceleration
 mechanism is hadronic and is responsible for the observed flux of cosmic
 rays.  The energy spectrum predicted by Waxman and Bahcall is shown in
 figure \ref{fig:intensities}.  A diffuse search for gamma-ray bursts involves
 summing over the observation time and spatial search windows for many
 separate bursts.

We considered at the moment an observation only over the northern sky, where the
search will not be limited to higher energies due to downgoing CR muon
background. A downgoing search would be similar, only it would require 
a harder energy cut.
From the remaining rate of 500 bursts per $2\pi$\,sr and year
we would expect 13 neutrino induced up-going muons at 0.1 
background events.  For this analysis we
used a hypothetical observation duration of 10 seconds and a spatial search
cone of $10^\circ$ centered about the direction of the GRB. The upper limit
to the associated GRB flux would be 20\% of the flux calculated by Waxmann
and Bahcall.

\section{ OTHER SCIENCE OPPORTUNITIES}

\subsection{ Dark matter }
If Weakly Interacting Massive Particles (WIMPs) make up the dark matter of
the universe, they would also populate the galactic halo of our own Galaxy. 
They would get captured by the Earth or the sun where they would annihilate
pairwise, producing high-energy muon neutrinos that can be searched for by
neutrino telescopes.  A favorite WIMP candidate is the lightest neutralino
which arises in the Minimal Supersymmetric Model (MSSM).  The typical
energy of the neutrino-induced muons would be of the order of $\approx 25\%$
of the neutralino mass.  The predicted muon rates from WIMPs annihilating
in the Sun range up to $10^4 km^{-2} $
per year at $\approx$100\,GeV and more 
than a $10^{4}km^{-2}$ up to 
energies of $\approx$1\,TeV. Current limits lie at fluxes of
several thousand events at energies above 100\,GeV. Simulations indicate
that IceCube should reach sensitivities below 50 muon events per year for
WIMPs from the sun \cite{edsjoe}. Analysis methods for WIMP searches 
be optimized for low energies. 
IceCube could play a complementary
role to future direct detection experiments (like CRESST or GENIUS) for
annihilation in the Earth, and may even have a slight advantage over direct
detection experiments for certain low-mass WIMP models and annihilation in
the Sun.

\subsection{  Cosmic rays and airshowers }

Combined with the 1\,km$^2$ surface detector IceCube can do unique
coincidence and anti-coincidence measurements with high energy air showers. 
In addition to providing a sample of events for calibration and for study
of air-shower-induced backgrounds in IceCube, the surface array will act as
a partial veto.  All events generated by showers with $E > 10^{15}$ eV can
be vetoed when the shower passes through the surface array.  In addition,
higher energy events, which are a potential source of background for
neutrino-induced cascades, can be vetoed even by showers passing a long
distance outside the array. 
The IceCube-IceTop coincidence data will cover the energy range from below
the knee of the cosmic-ray spectrum to $>10^{18}$eV. Each event will
contain a measure of the shower size at the surface and a signal from the
deep detector produced by muons with $E > 300\,GeV$ at production.  At the
high elevation of the South Pole, showers will be observed near maximum so
that measured shower size provides a good measure of the 
total shower size. 
The combined measurement of the muon-induced signal in IceCube 
 and the shower size at the surface will give a new
measure of primary composition over three orders of magnitude in energy. 
In particular, if the knee is due to a steepening of the rigidity spectrum,
a steepening of the spectrum of protons around $3\cdot 10^{15}$\,eV should
be followed by a break in the spectrum of iron at $8\cdot 10^{16} eV$.  The
method to measure the composition has been developed and applied
successfully with AMANDA and the surface airshower array SPASE-2
\cite{kath-thesis}.

\subsection{ SUPERNOVAE }

Although the MeV-level energies of supernova neutrinos are far below the AMANDA/IceCube
trigger threshold, a supernova could be detected by observing higher
counting rates of individual
PMTs over a time window of 5-10 s. The enhancement in rate of one PMT will be buried
in dark noise signals of that PMT. However, summing the signals from all PMTs over 10 s,
significant excesses can be observed. With background rates more than 10 times lower than
ocean experiments, IceCube has the potential to see a supernova and to generate
an alarm signal. 

\subsection{Cascades and very high energies}
Due to its large and regularly instrument volume, IceCube has a 
high sensitivity for cascades generated by 
 $\nu_e$, $\overline{\nu}_e$, $\nu_\tau$, and
$\overline{\nu}_\tau$.
Cascades have a superior energy resolution because all energy is 
deposited on a small volume of $\approx$10 metres diameter, and the 
amount of Cherenkov photons scales linearly with the deposited energy.
At energies above a few tens of TeV the detector becomes fully efficient
to cascade detection with an effective volume comparable to the 
geometric volume of 1\,km$^{3}$.
At energies above 1 PeV chances increase to distinguish tau events
from the  $\nu_e$-induced cascades. Extremely energetic muon 
events reach effective areas well beyond the 
geometric area as indicated in figure \ref{fig:aeff.en}. Effective 
areas at 10$^{18}$\,eV are estimated to reach more than 2.5\,km$^{2}$.
Reconstruction algorithms for the more complex event topologies 
such as from tau events and for extreme energies beyond 10$^{18}$\,eV 
are not developed yet. However, the high resolution and high dynamic range 
data acquisition system of IceCube will provide rich information to 
extract the physics from such events.

\section{ACKNOWLEDGEMENTS}

This material is based upon work supported by the following agencies: 
The U.S. National Science 
Foundation under Grant Nos. OPP-9980474  and OPP-0236449; 
University of Wisconsin Alumni Research
Foundation; U.S. Department of Energy; Swedish Research Council; Swedish Polar Research
Secretariat; Knut and Alice Wallenberg Foundation, Sweden; Federal
Ministry for Education and Research (Germany); 
U.S.  National Energy Research
Scientific Computing Center (supported by the Office of Energy
Research of the U.S.  Department of Energy); 
Deutsche Forschungsgemeinschaft (DFG).

\end{document}